\documentclass[prb,preprint,superscriptaddress,preprintnumbers,amssymb]{revtex4}
\usepackage{graphicx}
\usepackage{dcolumn}% Align table columns on decimal point
\usepackage{bm}% bold math
\usepackage{graphics}
\usepackage{braket}
\usepackage{siunitx}
\usepackage{nicefrac}
\usepackage[pdftex]{hyperref}
\usepackage{xcolor}
\usepackage{xspace}
\usepackage{comment}
\usepackage{float}
\usepackage{soul}
\usepackage{doi}
\usepackage[normalem]{ulem}
\usepackage{amsmath} 
\newcommand{\vinca}{\affiliation{Vin\v{c}a Institute of Nuclear Sciences - 
National Institute of the Republic of Serbia, University of Belgrade, P. O. Box 522, RS-11001 Belgrade, Serbia}}
\newcommand{\cnrspin}{\affiliation{Consiglio Nazionale delle Ricerche CNR-SPIN, c/o Universit\'{a} degli Studi "G. D’Annunzio", 66100 Chieti, Italy}}
\newcommand{\cnrrome}{\affiliation{Consiglio Nazionale delle Ricerche CNR-SPIN, Area della Ricerca di Tor Vergata, Via del Fosso del Cavaliere, 100, I-00133 Rome, Italy}}

\newcommand{\angstrom}{\mbox{\normalfont\AA}}

\newcommand{\cluster}{CrI$_n$--I$_n$Cr\xspace}

\begin{document}

\title{Delving into the anisotropic interlayer exchange in bilayer CrI$_3$}
%
% AUTHORS
\author{Srdjan Stavri\'c} \cnrspin \vinca \email{srdjan.stavric@spin.cnr.it}
\author{Paolo Barone} \cnrrome
\author{Silvia Picozzi} \cnrspin

\date{\today}
 
\begin{abstract}
\noindent
Bilayer CrI$_3$ attracted much attention owing to peculiar switching between the layered ferromagnetic and antiferromagnetic order upon stacking alternation. This finding pointed out the importance of the apparently small interlayer exchange, yet, existing literature addresses only its isotropic part. To fill this gap, we combine the density functional theory with Hamiltonian modeling to examine the anisotropic interlayer exchange in bilayer CrI$_3$ -- Dzyaloshinskii-Moriya (DMI) and the Kitaev interaction (KI). We develop and apply a novel computational procedure that yields the off-diagonal exchange matrix elements with $\mu{\rm eV}$ accuracy. Inspecting two types of bilayer stacking, we 
found a weak interlayer KI and much stronger DMI between the sublattices of monoclinic bilayer and their complete absence in rhombohedral bilayer. We show how these anisotropic interactions depend on the interlayer distance, stacking sequence, and the spin-orbit coupling 
strength and suggest the dominant superexchange processes at play. In addition, we demonstrate that the single-ion anisotropy largely depends on stacking, increasing by 50\% from monoclinic to rhombohedral structure. Remarkably, our findings prove that iodines, owing to their spatially extended $5p$ orbitals featuring strong spin-orbit coupling, are extremely efficient in mediating DMI across the van der Waals gap in two-dimensional magnetic heterostructures. Given that similar findings were previously demonstrated only in metallic multilayers where the DMI shows a much longer range, our study gives promise that the chiral control of spin textures can be achieved in two-dimensional semiconducting magnetic bilayers whose ligands feature strong spin-orbit coupling.

%\vskip 1cm
%\textbf{Keywords}: \textit{interlayer exchange}, \textit{bilayer CrI$_3$}, \textit{Dzyaloshinskii-Moriya interaction}, \textit{magnetic anisotropy energy}\\

\end{abstract}
\maketitle

\section{Introduction} 
\label{sec.intro}
%
% general introductory paragraph 
Since the long sought discovery of two-dimensional (2D) magnets finally happened with 
CrI$_3$ and Cr$_2$Ge$_2$Te$_6$ \cite{Huang2017Jun,Gong2017Jun}, a large stream of scientific 
efforts has been directed towards achieving new capabilities with magnetic van der Waals (VdW) heterostructures \cite{Klein2018Jun,Song2018Jun,Wang2018Jun,Gibertini2019May,Soriano2020Sep}. 
Given the diversity of 2D materials, there is a profusion of possible magnetic VdW heterostructures that offer endless possibilities. 
Yet, to find intriguing phenomena, one doesn't need to look any 
further from the two layers of CrI$_3$. So far, the magnetic properties of bilayer CrI$_3$ are 
manipulated by electric fields \cite{Huang2018Jul,Morell2019Feb}, electrostatic doping\cite{Jiang2019Apr}, pressure \cite{Song2019Dec}, and twisting \cite{Xu2022Feb}. In addition, theoretical studies proposed to switch the direction of magnetization in one of its layers by spin-orbit torque\cite{Dolui2020Mar} and predicted the magnetic photogalvanic effect\cite{Zhang2019Aug}, magnetic polarons \cite{Soriano2020Jan}, and magnetoelectric response in bilayer CrI$_3$\cite{Lei2021Mar}. 
Yet, to efficiently exploit the possibilities offered by bilayer CrI$_3$ in new concept devices, one has to truly understand the mechanism of the exchange coupling between the layers. 

% monolayer CrI3
Monolayer CrI$_3$ is composed of chromium atoms arranged in a honeycomb lattice surrounded by edge--sharing 
iodine octahedra. Below the Curie temperature the $S=3/2$ spins on Cr atoms are parallel and the monolayer CrI$_3$ is a ferromagnet (FM) with an out-of-plane magnetization \cite{Lado2017Jun, Besbes2019Mar, Kashin2020Mar}.
%The octahedral I$^{-}$ ligand field splits the Cr-$3d$ orbitals 
%into $t_{2g}$ and $e_g$ subsets. 
%Thus the CrI$_3$ is a semiconductor with a bandgap of $1.1-1.2 \, \rm{eV}$\cite{Yu2022Apr}. 
%The three lower $t_{2g}$ orbitals are half occupied by three spin-$\uparrow$ electrons 
%resulting with $S=3/2$ spin configuration of Cr ion.
The magnetic anisotropy energy (MAE), that is absolutely necessary for the 
long-range magnetic order to persist in 2D crystals at finite temperatures\cite{Mermin1968}, 
emerges in CrI$_3$ from an interplay between the single-ion anisotropy (SIA) 
and the two-ion anisotropy (TIA) %\sout{Kitaev interaction (KI)} 
occurring between the nearest neighbors Cr ions \cite{Lado2017Jun,Xu2018Nov}. The TIA is usually  derived within the generalized Heisenberg-Kitaev model and gives rise, in addition to the conventional isotropic Heisenberg exchange,  to the Kitaev exchange and to the symmetric pseudo-dipolar interaction, depending on the bond-orientation.
These terms, which we will generally label as  "Kitaev-like interaction" (KI), refer to the traceless symmetric part of the most general expression for bilinear spin-spin interactions\cite{Moriya1960Oct} and cooperate with dipole-dipole interaction in shaping the total magnetic anisotropy\cite{evans2020prb}. 
Like SIA, the KI in CrI$_3$ comes mostly from the spin-orbit coupling (SOC) on iodine atoms. MAE scales with the ligand SOC strength, being instrumental to CrI$_3$ showing a higher Curie temperature ($T_C$) than the isostructural CrBr$_3$ and CrCl$_3$, whose ligands feature much weaker SOC than iodines\cite{Webster2018Oct,Lu2019Nov,Xue2019Dec,Lu2020Sep}.  Besides SIA and KI, SOC can give rise to the Dzyaloshinskii-Moriya interaction (DMI). However, the presence of an inversion center in the nearest neighbor Cr-Cr bonds imposes that this antisymmetric part of the anisotropic exchange is exactly zero. On the other hand the DMI is allowed between the next-nearest Cr neighbors and, although being tiny, it can play an important role in gaping the magnon spectra\cite{Jaeschke-Ubiergo2021May}. 
%The Kitaev interaction is alias for the symmetric part of the anisotropic exchange and, like SIA, in monolayer CrI$_3$ it comes mostly from the spin-orbit coupling (SOC) on iodine atoms. Because of its SOC origin MAE scales with SOC strength, which explains why CrI$_3$ has higher Curie temperature than the isostructural CrBr$_3$ and CrCl$_3$ whose ligands feature much weaker SOC than iodines\cite{Webster2018Oct,Lu2019Nov,Xue2019Dec,Lu2020Sep}.

%This is a well known story told many times, yet still there are some doubts.
%For instance, the nearest-neighbor ferromagnetism of CrI$_3$ was usually attributed 
%to nearly $90^\circ$ Cr-I-Cr bond angle that favors the FM superexchange, 
%as predicted by the Goodenough-Kanamori (GK) rule \cite{Goodenough1958Aug,Kanamori1959Jul}. 
%However, recent theoretical studies suggest that the correct prediction of GK rule may be accidental 
%as it is not a single but a multiple superexchange processes that lead to the ferromagnetism of CrI$_3$\cite{Besbes2019Mar,Kashin2020Mar,Song2022Dec}.

%What is written so far overviews the very basics of the intralayer magnetism in monolayer CrI$_3$, 
%but the questions of merit for this work arise from the interlayer coupling in its bilayer.
In addition to magnetic properties that bilayer CrI$_3$ inherits from its constitutive layers, the interlayer exchange has proven extremely important as it can affect the direction of layers' magnetizations.    
It is an order of magnitude weaker than the intralayer exchange, but the possibility to tune it via stacking alternations made it a subject of numerous studies. 
For example, if we adopt the stacking from bulk CrI$_3$ that crystallizes either in rhombohedral (the low temperature or LT phase, $R\bar{3}$ space group) or monoclinic lattice
(the high temperature or HT phase, $C/2m$ space group) \cite{McGuire2015Jan}, 
we end up with two different bilayer structures that we refer to as the LT and the 
HT structure (Fig.~\ref{fig.stacking}a-b). 
Here the theoretical\cite{Sivadas2018Dec, Jang2019Mar, Jiang2019Apr, Soriano2019Sep, Kong2021Oct} and experimental\cite{Li2019Dec} studies agree: the LT stacking favors the FM ordering of layers' magnetizations, whereas the HT stacking leads to layered antiferromagnetic (AFM) order. 
%Therefore, despite the weakness of the VdW interlayer interaction, the overlap of the I-$5p$ orbitals 
%in the VdW gap of the bilayer is extremely important for its magnetism.
%It turns out that their magnetic properties are much different, as the overlap of the I-$5p$ orbitals in the VdW gap heavily depends on the stacking sequence.
However, being realized through (at least) two iodines, the interlayer Cr-Cr coupling is 
mostly of super-superexchange type, which makes its microscopic description complicated due to a high number of relevant hopping processes\cite{Ke2021Jan,Song2022Dec}.

In bilayer CrI$_3$ studies galore the interlayer exchange is most often considered isotropic. This is a reasonable assumption, given that the bilayer CrI$_3$ lattice is centrosymmetric and thus the DMI is forbidden, whereas the KI, if nonzero, is usually very small. However, the strict constraints imposed on DMI by the global symmetry of the structure by no means forbid the DMI to appear locally, between the specific neighbors.
Moreover, if the DMI exists between the specific neighbors and the global symmetry constraints are somehow removed -- the macroscopic DMI can emerge as well. 
New studies warm up such expectations showing that skyrmions -- topologically protected particle-like spin textures that usually appear as a consequence of DMI -- can be induced via moire magnetic exchange interactions in twisted bilayer CrI$_3$\cite{Akram2021Aug,Xu2022Feb,Yang2023Apr}. 
%To warm up the expectations, Sun \textit{et al.} reported the giant nonreciprocal second harmonic generation (SHG) in monoclinic CrI$_3$ bilayer\cite{Sun2019Aug}. In centrosymmetric crystals the SHG does not appear unless the spatial-inversion symmetry is broken by some other means. %Here, it is the layered AFM order that breaks both the spatial-inversion and the time-reversal symmetries, thus allowing SHG. 
%This experimental finding gives hope that the interlayer DMI can be induced by external manipulations, similar to that proposed for monolayer CrI$_3$\cite{Liu2018Feb, Behera2019Jun,Zhang2022Sep}. Previous studies the interlayer DMI is reported in synthetic antiferromagnets, proving that this antisymmetric interaction can be long-ranged if mediated through a proper spacer\cite{Fernandez-Pacheco2019Jul,Han2019Jul}. 
Speaking of interlayer DMI in general, Vedmedenko \textit{et al.}\cite{Vedmedenko2019Jun} proposed the atomistic model that 
predicts the formation of global chiral spin textures due to interlayer DMI
between the ferromagnetic layers coupled through a nonmagnetic spacer.
Recently, chiral control of spin textures is experimentally achieved in ferromagnetic TbFe/Pt/Co thin films, where the out-of-plane magnetization of TbFe is DMI-coupled with 
the in-plane magnetization of Co\cite{Avci2021Oct}. In this multilayer system the interlayer DMI is strong because the Pt atoms carry the conductive electrons that feature strong SOC. Having this in mind, the question is whether the iodine ligands, that also have considerable SOC, can play the role of DMI-mediator in bilayer CrI$_3$.
Moreover, having in mind the importance of KI in monolayer CrI$_3$, how strong is the interlayer KI in bilayer CrI$_3$?

We present a theoretical study that combines the density functional theory (DFT) and Hamiltonian modeling in calculating the anisotropic part of the interlayer exchange in bilayer CrI$_3$. The manuscript is organized as follows: we start by presenting the employed computational approach in Subsection~\ref{subsec.approach} and describe the model Hamiltonian that expresses the coupling between two perpendicular spins in Subsection~\ref{subsec.model}. The ability of the perpendicular-spins model to capture the changes in DFT band energies is demonstrated in Subsection~\ref{subsec.clusters}. The validity of the model is extended to describe the coupling between fully magnetized layers in Subsection~\ref{subsec.layers}, where we reveal a considerable DMI and an order of magnitude weaker KI between the sublattices of bilayer CrI$_3$. We demonstrate how both DMI and KI depend on structural transformations and the SOC strength in Subsection~\ref{subsec.dmi_ki}, suggesting possible superexchange mechanisms governing these interactions. Finally, in Section~\ref{sec.conclusion} we summarize the study by proposing a 2D magnetic heterostructure that should be a suitable platform for realizing the interlayer DMI coupling of layers' magnetizations, similar to that experimentally achieved in metallic thin films.

\section{Results}
\label{sec.results}
\subsection{Computational approach}
\label{subsec.approach}
\begin{figure}[h]
	\centering
	\graphicspath{ {figs/} }
	\includegraphics[width=1.0\linewidth]{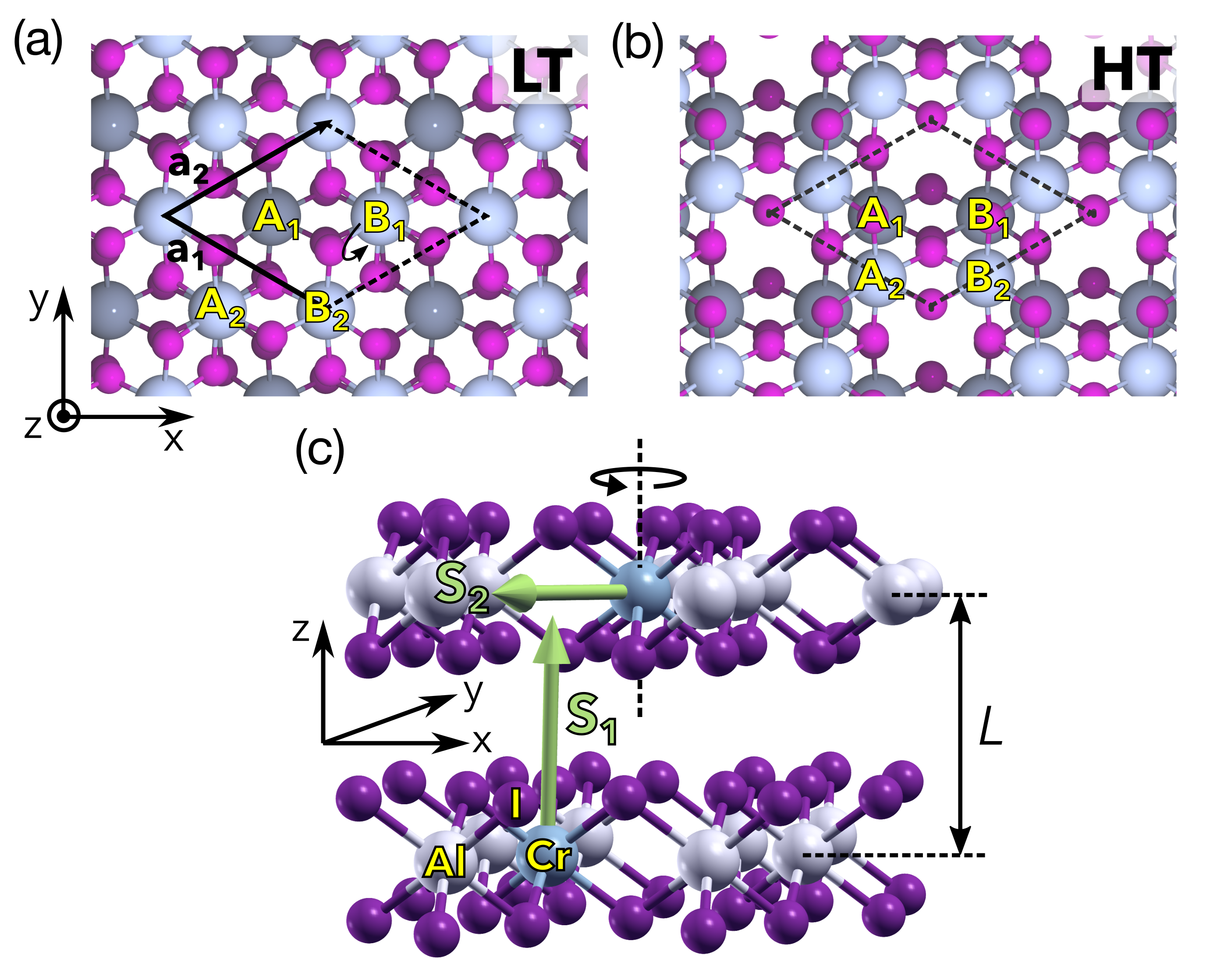}
	\caption {
	Bilayer CrI$_3$ with LT (a) and HT (b) stacking. Atoms from the upper layer are colored by brighter nuances. The HT structure is obtained from LT when the upper layer is translated by ${\bf a}_2/3$	
	(c) One pair of Cr atoms immersed into the CrI$_3$-like environment with Cr atoms replaced by Al. 
	Spins on Cr atoms are depicted by thick green arrows. The rotation axis is parallel to ${\bf S}_1$ and 
	the spin quantization axis is parallel to ${\bf S}_2$.  
	}
	\label{fig.stacking}
\end{figure}

\noindent
We model the bilayer CrI$_3$ by stacking two CrI$_3$ layers in rhombohedral (LT)
and monoclinic (HT) sequences (Fig.~\ref{fig.stacking}). The structural details and computational parameters
of DFT calculations are given in the Section \ref{sec.methods}. 
In order to study the coupling between the two spins from different layers, 
we need to isolate them from the rest of the magnetic environment. 
To solve this problem, we use the $2 \times 2$ supercell 
and replace by Al all the Cr atoms except the two inspected ones  (Fig.~\ref{fig.stacking}c). 
Al, like Cr, is trivalent so it doesn't perturb much the surrounding iodine ligand field. Therefore, the two remaining Cr atoms end up embedded into nonmagnetic crystalline environment that is reminiscent of that in bilayer CrI$_3$.  
To check the validity of the atomic replacement method, we calculated the SIA and the intralayer nearest-neighbor exchange tensor $\mathcal{J}_1$ of monolayer CrI$_3$ and compared the results to those obtained with the reference four-state method \cite{Xiang2012Dec} in Supplementary Information. 
%\footnote{NOTE FOR MYSELF: I used PBE alat and Cr-Cr distance from experiments as these calculations are done at the time when I didn't know any better. The difference in alat is 7.005 A (PBE) vs 6.874 A (PBE+VdW-DF2) but let's not complicate the story further with two different alats. I am confident that the results don't depend much on alat.} 
%
\par
In calculating the interlayer exchange tensors, we follow a two-step computational procedure: in the first step the electron density and the Kohn-Sham wavefunctions are obtained from noncollinear self-consistent field calculations (SCF) without SOC.
The directions of spins ${\bf S}_1$ and ${\bf S}_2$ on Cr atoms are 
constrained perpendicular to one another (Fig.~\ref{fig.stacking}c) 
using the penalty functional\cite{Ma2015Feb}. 
In the second step the Kohn-Sham wavefunctions from the first step are used, the SOC is included on all the atoms and the sum of the band energies are calculated for different directions of the spin quantization axis.  
The spin ${\bf S}_2$ is kept parallel to the spin quantization axis 
so that it rotates together with it, whereas the direction of ${\bf S}_1$ is unaffected 
by this rotation (see Fig.~\ref{fig.stacking}c). 
Given that only the sum of band energies is used, our procedure can be 
applied to any system provided that the use of the magnetic force theorem is justified\cite{Liechtenstein1987May,Solovyev2021Mar,Soriano2021Oct}. The benchmark of our method is provided in Supplementary Information, but here we want to stress that our method can save more than 50\% of computational time without affecting the accuracy of the reference four-state method \cite{Xiang2012Dec,Li2021Feb}. 

\subsection{Hamiltonian of two perpendicular spins}
\label{subsec.model}
\noindent
Here we derive the model describing the coupling of two perpendicular spins, that is 
used for mapping the differences between DFT band energies obtained for different directions 
of the spin quantization axis. In closer detail, we suppose that the two spins belong to different layers, 
but the model works perfectly fine if spins are from the same layer.
 
The total energy of such two-spin system is 
\begin{equation}
\label{eq.Etot}
E =  E_{\rm nm} + {\bf S}_1 \mathcal{J} {\bf S}_2 + {\bf S}_1 \mathcal{A}_1 {\bf S}_1 + {\bf S}_2 \mathcal{A}_2 {\bf S}_2,  
\end{equation}
where $E_{\rm nm}$ is the nonmagnetic contribution, $\mathcal{J}$ is an exchange tensor that couples 
spins ${\bf S}_1$ and ${\bf S}_2$, and $\mathcal{A}_i$ is the SIA tensor of layer $i$. 
If the layers are of the same material and magnetic atoms are Wyckoff partners sharing the same site symmetries,
%stacking preserves some symmetry relation between the layers
we have $\mathcal{A}_1 = \mathcal{A}_2 = \mathcal{A}$.
Without loss of generality let us choose the coordinate system by fixing ${\bf S}_1$ along the $z$-axis and 
placing ${\bf S}_2$ in the $xy$-plane. Now, ${\bf S}_1 = (0,0,S_1)$ and ${\bf S}_2 = (S_2 \cos(\phi), S_2 \sin(\phi), 0)$ so that the spin configuration is completely determined by a single parameter -- the angle $\phi$ between the ${\bf S}_2$ and the $x$-axis. The Eq.~\ref{eq.Etot} turns into
\begin{equation}
\begin{split}
	E(\phi) &= E_C + S_1 S_2 \Big( J_{zx} \cos(\phi) + J_{zy} \sin(\phi) \Big) + S_2^2 \Big( A_- \sin^2(\phi) + A_+ \sin(2\phi) \Big) \\ 
	&= E_C + E_J(\phi) + E_A(\phi),
\end{split}
\label{eq.Ephi}
\end{equation}
where we introduced the parameters $A_- \equiv A_{yy} - A_{xx}$ and $A_+ \equiv (A_{xy}+A_{yx})/2$ for convenience and $E_C = E_{\rm nm} + A_{zz}S_1^2 + A_{yy}S_2^2$ comprises all the terms that are $\phi$-independent. 
The important terms for our discussion are the exchange interaction energy $E_J(\phi)$ and the contribution from the 
single-ion anisotropy $E_A(\phi)$. 
If we interchange the roles of ${\bf S}_1$ and ${\bf S}_2$ the Eq.~\ref{eq.Ephi} yields 
$J_{xz}$ and $J_{yz}$.
If we further change the rotation axis from $z$ to $x$ we obtain $J_{xy}$ and $J_{yx}$ 
(and $J_{xz}$ and $J_{zx}$ that we have already). 
Therefore, by considering the sets of spin configurations that we denote as $\big\{ {\bf S}_1 \parallel z \leftrightarrow {\bf S}_2 \circlearrowleft z \big\}$ and $\big\{ {\bf S}_1 \parallel x \leftrightarrow {\bf S}_2 \circlearrowleft x \big\}$, one obtains all the \textit{off-diagonal} $\mathcal{J}$-matrix elements.
%\footnote{$\big\{ {\bf S}_1 \parallel z \leftrightarrow {\bf S}_2 \circlearrowleft z \big\}$ denotes all the spin configurations that are obtained from fixing ${\bf S}_1$ along the $z$-axis and rotating the ${\bf S}_2$ around it. $\leftrightarrow$ denotes that the configurations with the interchanged roles of ${\bf S}_1$ and ${\bf S}_2$ are included as well.}

\par
In general, the $\mathcal{J}$ tensor decomposes into the isotropic Heisenberg exchange and the anisotropic DMI (antisymmetric) and KI (symmetric)\cite{Li2021Feb},
\begin{equation}
		\mathcal{J} = \underset{\rm Heisenberg \; exchange}{\frac{1}{3} {\rm Tr}(\mathcal{J}) \mathbb{I}_3} + 
	\underset{\rm DMI}{\frac{1}{2} (\mathcal{J} - \mathcal{J}^T)} + 
	\underset{\rm anisotropic\; symmetric \; exchange}{\Big[\frac{1}{2} (\mathcal{J} + \mathcal{J}^T)- \frac{1}{3} {\rm Tr}(\mathcal{J})\mathbb{I}_3 \Big]}
	= \frac{1}{3} {\rm Tr}(\mathcal{J}) \mathbb{I}_3 + \mathcal{D} + \mathcal{K}.
	\label{eq.J_decomposition}
\end{equation} 
By assumption, the spins are perpendicular and the Heisenberg exchange between them vanishes. 
Thus, the exchange energy contains only the DMI and the KI contributions, 
$E_J(\phi) = E_{\rm DM}(\phi) + E_K(\phi)$. The DMI is usually expressed as the mixed vector product,
$E_{\rm DM} = {\bf D} \cdot ( {\bf S}_1 \times {\bf S}_2 )$, which introduces the \textit{Dzyaloshinskii vector},
\begin{equation}
	{\bf D} = (D_x, D_y, D_z) = \frac{1}{2} \Big( J_{yz}-J_{zy}, J_{zx}-J_{xz}, J_{xy}-J_{yx} \Big).
\end{equation} 
%
%The magnitude of DM vector $|{\bf D}| = D $, usually displayed in the units of energy, determines the strength of DM interaction. 
For a system with a well defined symmetry, the direction of ${\bf D}$ can be (at least partially) determined with 
the help of Moriya rules\cite{Moriya1960Oct}. In Supplementary Information we derive the Moriya rules in a form that is more suitable for our purposes. 

\par
Within $\big\{ {\bf S}_1 \parallel z \leftrightarrow {\bf S}_2 \circlearrowleft z \big\}$ spin configurations, 
the total variation of the Kitaev energy is $\Delta E_K = 2 S_1 S_2 \big( K_{zx}^2 + K_{zy}^2 \big)^{1/2}$. 
In analogy to DMI, it is convenient to introduce the vector ${\bf K} = \frac{1}{2} \big( J_{yz}+J_{zy}, J_{zx}+J_{xz}, J_{xy}+J_{yx} \big)$ that quantifies the strength of KI by $\Delta E_K / 2 S_1 S_2 = (K_x^2 + K_y^2)^{1/2}$. Note that in general the matrix $\mathcal{K}$, unlike $\mathcal{D}$, contains 
the diagonal elements as well that are out of reach of this method as 
their calculation require the (anti)parallel spin configurations. 
Therefore, strictly speaking for a given reference frame the parameter $K$ quantifies the strength of the off-diagonal part of the Kitaev interaction. 
%\footnote{We must be cautious when referring to the Kitaev interaction and should always keep 
%in mind that our model assumes the spins are perpendicular so that we miss the contributions from diagonal
%$\mathcal{K}$-matrix elements.} 
%Hence, if one has all the off-diagonal $\mathcal{J}$-matrix elements then the DM vector is known as well. 

\par
From  Eq.~\ref{eq.Ephi}, we define the SIA constant as $A = \Delta E_A/S^2$, where $\Delta E_A$ is the total variation of the SIA energy term $E_A(\phi)$ for the rotation of spin ${\bf S}$ around a given axis.
%\footnote{Strictly speaking $A$ depends on the rotation axis and we should use a more precise 
%notation such as $A_{\bf n}$, where ${\bf n}$ is the rotation axis.} 
It can be easily derived that, for a fixed rotation axis, the energy $E_A(\phi)$ 
reaches extrema at $\phi_0^{\pm} = \arcsin \Big(\frac{1}{2} \pm \frac{A_+}{2\sqrt{A_+^2 + 4A_-^2}}\Big)^{1/2}$ and its total variation is $\Delta E_A = | E(\phi_0) - E(\phi_0 + \pi/2)|$.
%\footnote{$f(\phi) = a \sin^2(\phi) + b \sin(2 \phi)$ is a periodic function with periodicity $\pi$. Thus, if $E_A(\phi_0)$ is maximum then $E_A(\phi_0 + \pi/2)$ is minimum and vice versa. The absolute value signs are introduced not to worry about what is maximum and what is minimum.}. 
If spin rotates around the $z$-axis, the in-plane SIA constant $A_{\parallel}$ is obtained, whereas if 
it rotates around the $x$-(or $y$-) axis the fitting gives the out-of-plane SIA constant, $A_{\perp}$. 
%This and the calculations of DM and Kitaev constants for a CrI$_3$ bilayer we discuss in the next section. 
To sum up all being written about this model, if one performs the DFT calculations for $\big\{ {\bf S}_1 \parallel z \leftrightarrow {\bf S}_2 \circlearrowleft z \big\}$ and $\big\{ {\bf S}_1 \parallel x \leftrightarrow {\bf S}_2 \circlearrowleft x \big\}$
sets of spin configurations, the ${\bf D}$, ${\bf K}$, $A_\perp$, and $A_\parallel$ are obtained from fitting to Eq.~\ref{eq.Ephi}. We discuss this below for the cases of two Cr atoms from different layers.  

\begin{comment}
We can separate the exchange interaction energy from SIA energy term and quantify the relevant \textit{energy scales} for our problem. The first energy scale, defined by the difference between the maximum and the minimum of $E_J(\phi)$ divided by the magnitudes of the magnetic moments, is easily found to be $\Delta E_J / S_1 S_2 = 2 (J_{zx}^2 + J_{zy}^2)^{1/2}$. The second is $\Delta E_A / S_2^2$, where $ \Delta E_A = |E_A(\phi_0)-E_A(\phi_0+\pi/4)|$ and $\phi_0^{\pm} = \arcsin \Big(\frac{1}{2} \pm \frac{A_+}{2\sqrt{A_+^2 + 4A_-^2}}\Big)^{1/2}$ is the zero of $E_A'(\phi)$. \footnote{$f(\phi) = a \sin^2(\phi) + b \sin(2 \phi)$ is a periodic function with periodicity $\pi$. Thus, if $\phi_0$ is the maximum then $\phi_0 + \pi/4$ is the minimum and vice versa.} These two quantities will be used as main descriptors of the strength of interlayer exchange coupling and in-plane SIA induced in one layer by the presence of the other layer. 
\end{comment}

\subsection{Interlayer coupling of perpendicular Cr magnetic moments}
\label{subsec.clusters}
\begin{figure}[h]
	\centering
	\graphicspath{ {figs/} }
	\includegraphics[width=1.0\linewidth]{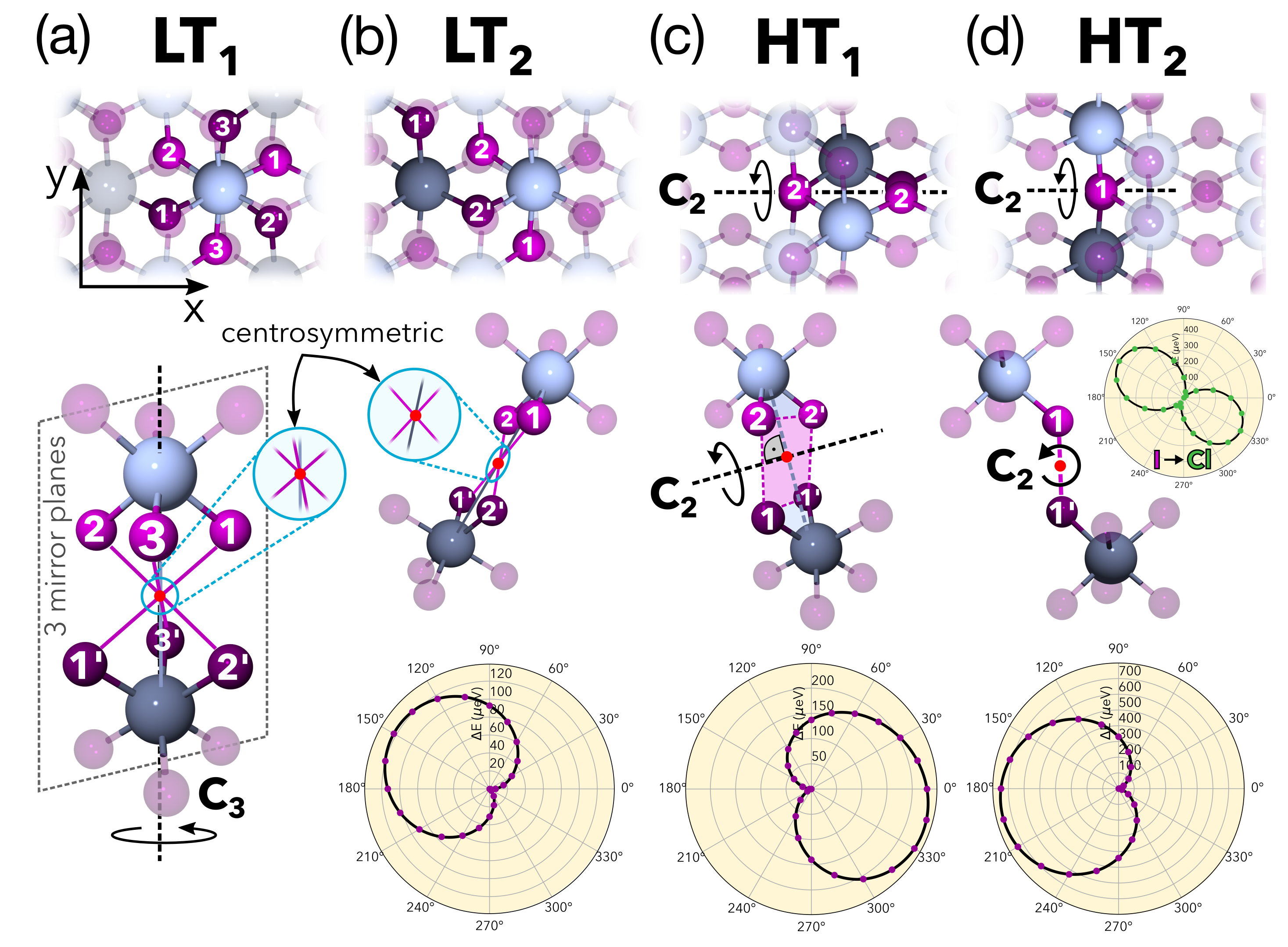}
	\caption {
		The structures used for modeling the nearest and the next-nearest interlayer Cr neighbors in LT and HT bilayer (upper panel). The four corresponding CrI$_n$-I$_n$Cr clusters are depicted in the middle panel and their symmetries are denoted. Yellow discs show the polar plots of the $E(\phi)- E_0$, where $E(\phi)$ is the sum of  band energies for the angle $\phi$ of the spin quantization axis, and $E_0$ is the minimum of band energies around the given rotation axis. In d) the $\infty$-shaped energy curve corresponds to the 
        HT$_2^*$ structure with I atoms from CrI--ICr cluster replaced by Cl.     
	}
	\label{fig.CrI3bl}
\end{figure}

\noindent
Bilayer CrI$_3$ is an illustrative example as it reveals a few different scenarios 
of interlayer exchange coupling and elucidates the important role of the symmetry 
of local environment around Cr ions. We will refer to Cr ions pair with its surrounding ligands as 
``the \cluster cluster" ($n$ is the number of ligands that participate in the exchange 
path between Cr ions). We consider the nearest and the next-nearest interlayer neighbors in both LT and HT structures. The coordinate system is chosen in a way so that the Cr-Cr bond is along the $x$-axis, while the $z$-axis is perpendicular to the CrI$_3$ plane (see Fig.~\ref{fig.CrI3bl}a).
For each considered structure, we perform the symmetry analysis to double check 
whether the calculated off-diagonal $\mathcal{J}$-matrix elements comply 
with the extended Moriya rules exposed in Supplementary Information. 

Starting with the LT bilayer, the nearest and the next-nearest interlayer Cr neighbors are modeled by LT$_1$ and LT$_2$ structures (see Fig.~\ref{fig.CrI3bl}a-b). In LT$_1$ the Cr-Cr bond displays inversion, threefold rotation axis parallel to the bond, and three vertical mirror planes. Therefore, the symmetry of CrI$_3$--I$_3$Cr 
cluster in LT$_1$ satisfies three Moriya rules (a,c,e) and consequently the exchange matrix must be diagonal. 
% Hence, the system does not display any anisotropic interaction.
In complete agreement our calculations give all the zeros at the off-diagonal slots of the exchange matrix ($\mathcal{J}_{\rm LT1}$ in Eq.~S1).
%\footnote{Based on the symmetry of the system, one can reduce the number of calculations needed to obtain the off-diagonal $\mathcal{J}$ matrix elements. This particular case is the most fortunate one as there is no need to perform calculations at all. Nonetheless, we did perform the calculations and indeed obtained all the zeros at the off-diagonal sites.} 
%
Moving further to next-nearest neighbors, the CrI$_2$--I$_2$Cr cluster in LT$_2$ has only the spatial-inversion symmetry (Moriya rule a) and 
the exchange matrix is symmetric, thus forbidding the DMI. On the other hand, 
no symmetry rule forbids the KI and our calculations reveal $K = 32 \, \mu{\rm eV}$ (Table~\ref{tab.dEJ_dEA}). 
From this example one can actually see the Kitaev interaction in action: with ${\bf S}_1$ fixed along the $z$-axis and ${\bf S}_2$ rotating around it, 
the energy of the system is changing according to the \textit{cardioid} pattern which is the direct consequence of the KI (see the polar plot in Fig.~\ref{fig.CrI3bl}b). 
%Here this is completely due to KI because DMI is forbidden, but the cardioid shape of the energy dependence can also be the consequence of DMI (or of both).  
%
%Notice that the energy dependence on the ${\bf S}_2$ direction in the polar plot in Fig.~\ref{fig.CrI3bl}b has a \textit{cardioid} shape. The twofold symmetry of cardioid is a consequence of the existence of preferred orientation for ${\bf S}_1 \times {\bf S}_2$, i.e. the two magnetic moments are chirally coupled.  
%Note that in CrI$_3$ bilayer the Cr1-Cr2 coupling can be characterized as a super-super-exchange (the "interaction path" passes through the two iodines). Hence, the DMI is expected to be much weaker than that stemming from the super-exchange. 
%
\begin{table}[h]
	\caption{The magnitude and the components of Dzyaloshinskii and Kitaev vectors. 
		The last two columns present the out-of-plane and the in-plane SIA constants.}
	\begin{tabular}{ l||c|c|c|c||c|c|c|c||c|c} 
		\hline
structure $| \mu{\rm eV} \rightarrow$ & $|{\bf D}|$ & $D_x$ & $D_y$ & $D_z$ & $|{\bf K}|$ & $K_x$ & $K_y$ & $K_z$ & $A_{\perp}$ & $A_{\parallel}$ \\ \hline
		LT$_1$ (Fig.~\ref{fig.CrI3bl}a)  &  0  &  0    &   0   &    0  & 0 &   0    &  0   &   0   &  253        &  0              \\ 
		LT$_2$ (Fig.~\ref{fig.CrI3bl}b)  &  0  &  0    &   0   &    0  & 32&  -14   & 24   & -16   &  240        &  1              \\  
		HT$_1$ (Fig.~\ref{fig.CrI3bl}c)  & 175 &  0    &  50   & -168  & 0 &   0    &  0   &   0   &  172        & 19              \\
%		4 HT$_1$ (2-$1'$ Cl)               & 204 & -15   & 105   & 174   & 28&   9    & 15   &  22   &  275        & 190             \\ \hline
		HT$_2$ (Fig.~\ref{fig.CrI3bl}d)  & 166 &  0    & -166  &  -4   & 25&  -25   &  0   &   0   &  173        & 12              \\ 
		HT$_2^*$ (CrCl-ClCr)                   & 14  &  0    & -14   &   3   & 1 &  -1    &  0   &   0   &  163        & 186             \\ 
%		7 HT$_2$ (I-Cl)                    &     &       &       &       &   &        &      &       &             &                 \\
		\hline
	\end{tabular}
    \label{tab.dEJ_dEA}
\end{table}
\par
Moving the discussion to HT stacking, the $x$-axis is a twofold rotational axis 
for both CrI$_2$--I$_2$Cr in HT$_1$ and CrI--ICr cluster in HT$_2$ structure (see Fig.~\ref{fig.CrI3bl}c-d). 
Given that the Cr-Cr bond is perpendicular to $x$-axis, these two cases fall into the Moriya rule d category.
Taking into account the choice of the global coordinate system, 
the symmetry implies $J_{yx} = -J_{xy}$, $J_{zx} = -J_{xz}$, and $J_{zy} = J_{yz}$. 
%Further, in HT stacking nothing forbids the DMI between the nearest and the next-nearest neighbors and 
%by symmetry rules the ${\bf D}$ vector lies in the $yz$-plane. 
Although KI is not forbidden by symmetry, our calculations reveal 
its total absence  in HT$_1$ structure and a small $K$ of $25 \, \mu{\rm eV}$ in HT$_2$.

Strikingly, both the nearest and the next-nearest interlayer neighbors in HT structure display considerable DMI in the range of $150-200 \, \mu{\rm eV}$ (Table~\ref{tab.dEJ_dEA}). In Moriya's seminal paper, 
the DMI is derived from SOC and it is shown that the DM energy is linear in SOC\cite{Moriya1960Oct}. 
%We will thoroughly discuss the linearity of DMI in Subsection~\ref{subsec.dmi_ki}, but here we want to 
%stress the importance of ligands' SOC for DMI and KI. 
Given that the SOC on iodine (and not on chromium) gives the major contribution to MAE of monolayer CrI$_3$ \cite{Lado2017Jun,Xu2018Nov,Kim2020Feb}, we assume that it is also responsible for the interlayer DMI in bilayer CrI$_3$. This assumption can be tested through ligand replacement. 
The SOC constant for valence electrons in solids scales as $\lambda \sim 1/Z^2$, where $Z$ is the atomic number \cite{Shanavas2014Oct}. Therefore, if iodines are replaced with chlorine, by the rule of thumb the DMI would be reduced by $Z^2_{\rm I}/Z^2_{\rm Cl} \approx 10$ times. This simple estimate works surprisingly well, as we reveal that the $D$ is reduced 12 times when the two iodines in HT$_2$ cluster are replaced by chlorine (see Fig.~\ref{fig.CrI3bl}d and Table~\ref{tab.dEJ_dEA}). Moreover, this example proves the local character of the interlayer DMI, showing that 
only the ligands in the vicinity of Cr-Cr pair play a role in mediating this anisotropic interaction. Otherwise, the DMI would not be reduced so drastically even after the I$\rightarrow$Cl replacement in the CrI--ICr cluster because the other iodines 
are still present in the structure. The dependence of DMI on SOC is further discussed in Subsection~\ref{subsec.dmi_ki}.

\par
To the best of our knowledge, the effect of stacking on SIA is not addressed in previous studies of bilayer or bulk  CrI$_3$. Most often, it is assumed that the SIA obtained from calculations on the monolayer persists in bilayer and bulk. 
However, SIA is a local property and is thus sensitive to changes of the spin environment, e.g. by altering the stacking sequence in bilayer.
Remarkably, we obtained the increase of 50\% in $A_\perp$ from HT to LT stacking (Table~\ref{tab.dEJ_dEA}). 
Given that SIA largely contributes to magnetic anisotropy, one must take this change into account when estimating the critical temperature of bilayer or bulk CrI$_3$.   
Further, we obtained that LT bilayer is isotropic to the in-plane spin rotations ($A_\parallel = 0$) like in the monolayer CrI$_3$\cite{Lu2020Sep}, but not in 
HT bilayer where a small in-plane SIA of $A_\parallel \sim 10-20 \, \mu{\rm eV}$ is induced by stacking.  
This in-plane SIA is responsible for the distortion of the cardioid that corresponds to the HT$_1$ structure, Fig.~\ref{fig.CrI3bl}c.
%Although it is an order of magnitude smaller than $A_\perp$, the influence of the in-plane SIA can 
%be grasped from Fig.~\ref{fig.CrI3bl}c, where it is responsible for the cardioid distortion. 
Taking into account all being written in this Section, one should be truly careful in ascribing the monolayer magnetic properties to bilayer and bulk VdW magnets, as the interlayer interactions proved more important than was initially expected. 

\subsection{Interlayer coupling of perpendicular magnetizations}
\label{subsec.layers}
%
%In Subsection~\ref{subsec.clusters} we described the coupling of individual mutually perpendicular 
%Cr spins that belong to different layers. We showed that the exchange tensor can be calculated by mapping 
%the DFT energies to a simple two-spin model hamiltonian. 
%This brings us one step closer to our main goal, which is describing the coupling of the
%mutually perpendicular magnetizations (i.e. fully magnetized layers). 
In this Subsection we move away from the two-spin systems and provide the description of coupling 
between the fully magnetized layers with perpendicular magnetizations. With fully magnetized layers,
we assume that all the spins in the layer point to the same direction.
In general, when spins are randomly distributed, the total energy of $N \times N$ bilayer CrI$_3$ 
(that contains $2N^2$ spins per layer) is a sum of the nonmagnetic energy, 
the contributions from the \textit{intralayer} and \textit{interlayer} exchange coupling of spins, 
and SIA contributions at each spin site,  
\begin{equation}
	E_N = E_{N,\rm nm} + \sum_{i_1,i_2 = 1}^{2N^2} {\bf S}_{i_1} \mathcal{J}^{\updownarrow}_{i_1 i_2} {\bf S}_{i_2} 
	+ \sum_{l=1,2} \bigg[ \frac{1}{2} \sum_{i_l,j_l=1}^{2N^2} {\bf S}_{i_l} \mathcal{J}^{\leftrightarrow}_{i_l j_l} {\bf S}_{j_l} 
	+ \sum_{i_l=1}^{2N^2} {\bf S}_{i_l} \mathcal{A} {\bf S}_{i_l} \bigg],
	\label{eq.E_full_layer}
\end{equation}
where $l=1,2$ is the layer index and $i_l$ and $j_l$ are indices numbering the spin sites. The \textit{interlayer} exchange tensor $\mathcal{J}^{\updownarrow}_{i_1 i_2}$ describes the coupling between the spins $i_1$ and 
$i_2$ from different layers, whereas the \textit{intralayer} exchange tensor $\mathcal{J}^{\leftrightarrow}_{i_l j_l}$ describes the coupling between spins $i_l$ and $j_l$ from the same layer $l$. 

If we assume that all the spins belonging to one layer point to the same direction, 
i.e. ${\bf S}_{i_l} \equiv {\bf S}_l$, the Eq.~\ref{eq.E_full_layer} greatly simplifies,  
\begin{equation}
	E =  E_{\rm nm} + {\bf S}_1 \Big( \underbrace{\mathcal{J}^\updownarrow_{A_1 A_2} + \mathcal{J}^\updownarrow_{A_1 B_2} + \mathcal{J}^\updownarrow_{B_1 A_2} + \mathcal{J}^\updownarrow_{B_1 B_2}}_{\mathcal{J}^\updownarrow} \Big) {\bf S}_2 +
	\sum_l {\bf S}_l \Big( \frac{1}{2}\mathcal{J}^\leftrightarrow_{A} + \frac{1}{2}\mathcal{J}^\leftrightarrow_{B}
	+ 2 \mathcal{A} \Big) {\bf S}_l,
\label{eq.E_full_layer_simple}
\end{equation}
where $E = E_N/N^2$ and $E_{\rm nm} = E_{N, \rm nm}/N^2$ are the total and the non-magnetic energy expressed per unit cell.
In Eq.~\ref{eq.E_full_layer_simple} the contributions from $A$ and $B$ sublattices (see Fig.~\ref{fig.stacking}a,b) are separated in a sense that $\mathcal{J}^\updownarrow_{A_1B_2}$ describes the interlayer coupling between 
all the spins at $A_1$ sites with all the spins at $B_2$ sites ($N^2$ of each).
The tensor $\mathcal{J}^\leftrightarrow_{A}$ describes the interaction of a single spin at $A$ site with all the other spins from the same layer. If we specify the spins 
${\bf S}_1 = (0,0,S_1)$ and ${\bf S}_2 = (S_2 \cos(\phi), S_2 \sin(\phi), 0)$ the 
Eq.~\ref{eq.E_full_layer_simple} reads
\begin{equation}
	E(\phi) = E_C + S_1 S_2 \Big( J^\updownarrow_{zx} \cos(\phi) + J^\updownarrow_{zy} \sin(\phi) \Big) + S_2^2 \Big( Q_- \sin^2(\phi) + Q_+ \sin(2\phi) \Big), 
\end{equation}
which is the same equation as Eq.~\ref{eq.Ephi}. The parameters $Q_-$ and $Q_+$ stemming from $\mathcal{J}^{\leftrightarrow}$ and $\mathcal{A}$ will be used solely for fitting purposes and will not be further discussed, as our focus is on $\mathcal{J}^\updownarrow$.
%\footnote{This is to say that $Q_-$ and $Q_+$ will be used solely for fitting purposes and we will not try to extract the intralayer interaction and SIA contributions from them. Concretely, for fitting purposes, the most important is that the basis functions are linearly independent (as $\sin(x)$, $\cos(x)$, $\sin(2x)$ and $\sin^2(x)$ are) and that the interlayer terms do not mix with intralayer terms and SIA.}

\par  
We calculated the off-diagonal elements of $\mathcal{J}^\updownarrow$-matrix for HT bilayer 
and obtained a weak interlayer KI of $K = 18 \, \mu{\rm eV}$. 
On the other hand, already from the spatial-inversion symmetry of HT bilayer
we know that there is no DMI as the $A_1 A_2$ ($A_1 B_2$) contribution 
to the Dzyaloshinskii vector exactly cancels that from $B_1 B_2$ ($B_1 A_2$). 
%To reveal why is it so, we explicitly decompose the $\mathcal{J}^\updownarrow$ onto sublattices' contributions. 
Nevertheless, this does not mean that the interlayer DMI between the sublattices is zero.
To inspect this, we modeled the $A_1 A_2$ sublattice of the HT structure by replacing the 
Cr atoms of the $B_1 B_2$ sublattice with Al (Fig.~\ref{fig.DMI_KI}a) and obtained the 
Dzyaloshinskii vector with a magnitude of $D_{A_1 A_2} = 236 \, \mu{\rm eV}$.
This is a remarkable result, as the estimated $\frac{D}{|J|}$ ratio is 
80\% ($J$ from Ref. \cite{Jang2019Mar}), much higher than the 10\% threshold which 
is already considered promising for skyrmionics\cite{Koshibae2014Oct,Zhang2020Dec,Xu2020Feb}.
Compared with the experimental results, 
the $D_{A_1 A_2}$ is twice higher than the interlayer DM energy reported 
for TbFe/Pt/Co multilayers, which is among the strongest interlayer DMI 
realized in experiments\cite{Avci2021Oct}. 
% $D_{A_1 A_2} = 5.55 \, \mu{\rm eV}/\angstrom^2$ compared to Avci result 
% $44 \, \mu{\rm J}/{\rm m}^2 = 2.75 \, \mu{\rm eV}/\angstrom^2$ . 

\begin{comment}
Note that we could've obtained the DM vector between $A$ and $B$ sublattices by accounting the contributions from all 
the first and second nearest neighbors that we already calculated. 
\footnote{For the sake of completeness, we calculated the $J$-tensors between the third neighbors in HT 
structure using the $3 \times 3$ supercell. The DM of $D = 7 \, \mu{\rm eV}$ is clearly dwarfed by 
those of the first and the second neighbors.}. However, for some of the contributions we need to rotate by $120^\circ$
the $\mathcal{J}_{\rm HT1}$ and $\mathcal{J}_{\rm HT1}$ tensors in $xy$-plane, which we cannot do 
as we don't know the diagonal matrix elements.\footnote{One can also go the other way around and calculate
$\mathcal{J}_{\rm HT1}$ tensors for two Cr interlayer couples that are differently oriented 
and then reconstruct the diagonal elements by knowing that the two tensors must be connected by rotation. 
However, this is not computationally cheaper than the approach exposed in this section.}
\end{comment}

\subsection{Dependence of DMI/KI on structural transformations and SOC}
\label{subsec.dmi_ki}

\begin{figure}[h!]
	\centering
	\graphicspath{ {figs/} }
	\includegraphics[width=0.95\linewidth]{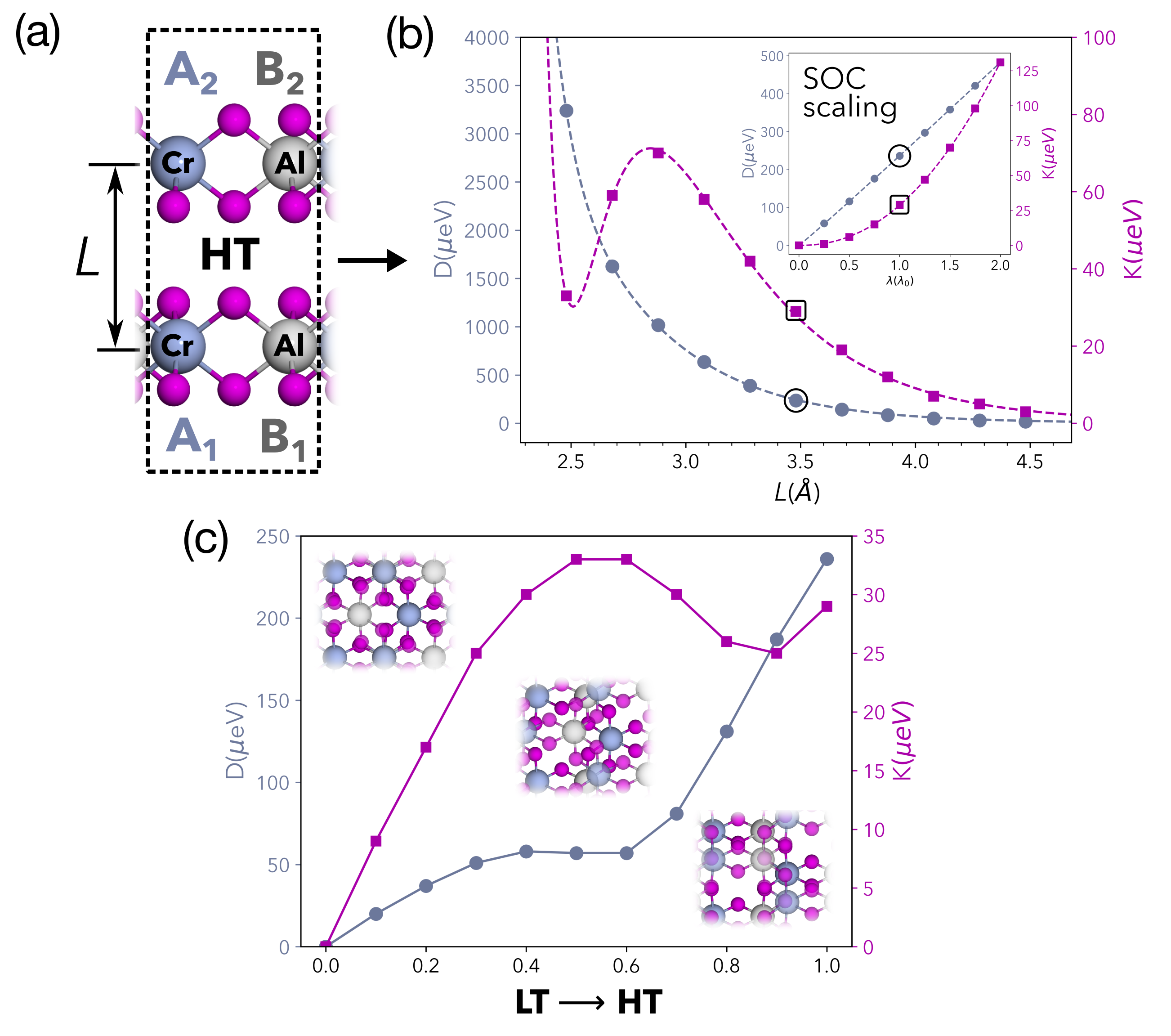}
	\caption {
		(a) Magnitudes of the Dzyaloshinskii $(D)$ and Kitaev $(K)$ vectors as a function 
of interlayer distance for $A_1 A_2$ sublattice of HT structure. Dashed lines are fitted lines with parameters 
$c_{10}^D=95~$~eV$\angstrom^{-10}$, $c_{10}^K=13~$~eV$\angstrom^{-10}$, $c_{14}^D=-5.46\times 10^3~$~eV$\angstrom^{-14}$, $c_{10}^K=-0.98\times 10^3~$~eV$\angstrom^{-14}$, $c_{18}^D=1.11\times 10^5~$~eV$\angstrom^{-18}$ and $c_{18}^K=0.19\times 10^5~$~eV$\angstrom^{-18}$.
 In the inset the $D$ and $K$ 
dependence on SOC constant is depicted as calculated for $L = 3.48 \, \angstrom$ (dashed lines displaying the fitting power function discussed in the text).
		(b) The $D$ and $K$ dependence on the stacking sequence from LT to HT structure for $L = 3.48 \, \angstrom$. Continuous lines are guides for the eye.
	}
	\label{fig.DMI_KI}
\end{figure}

As shown in Fig. \ref{fig.DMI_KI}b, the magnitude of both the DMI and KI display a fast, exponential-like decrease when the interlayer distance $L$ is larger than $\approx 3\,\angstrom$, consistently with the expected weak interaction between the layers across the VdW gap. At shorter separation, the evolution with $L$ suggests a more complicated situation, especially for the KI which displays a non-monotonic behavior. 
Assuming a superexchange-like mechanism for both interactions, the dependence on the interlayer distance can be rationalized taking into account the possible exchange paths involving different transfer integrals (hopping terms) between Cr and I ions and their expected dependence on the relative bond lengths $l$. One can identify three such hopping terms, corresponding to $d_{\text Cr}$-$d_{\text Cr}$ direct hopping ($t_1\propto l^{-5}$), $d_{\text Cr}$-$p_{I}$ hopping ($t_2\propto l^{-7/2}$) and $p_{\text I}$-$p_{\text I}$ hopping ($t_3\propto l^{-2}$), where we adopted the bond-length dependence of Harrison\cite{harrison_book}. In perturbation theory, the exchange interaction can be expressed quite generally as a sum of terms each with the form $t^n F(\lambda, U, \Delta)$, where $F(\lambda, U, \Delta)$ is a polynomial function of atomic SOC $\lambda$, charge transfer energy $\Delta$, and on-site Coulomb interaction $U$ and the exponent $n$ depends on the number of hopping terms involved in the exchange path. For instance, the contribution to the exchange interaction arising from the direct $d$-$d$ hopping between Cr atoms scales as $t_1^2 \propto l^{-10} \sim L^{-10}$. Similarly, a Cr-I-Cr exchange path involving a single ligand I--$p$ intermediate state would scale as $t_2^4 \sim L^{-14}$, while a Cr-I-I-Cr exchange path would also include a $p$-$p$ transfer integral, scaling overall as $t_2^4 t_3^2 \sim L^{-18}$. Neglecting the details of the bond geometries and the angular dependence of each transfer integral and assuming that the dependence on the interlayer distance inherits the same scaling law of hopping terms, one can derive a general functional expression for both $D$ and $K$ that reads
\begin{equation}
\label{eq:d-dependence}
I(L) =  \frac{c^I_{10}}{L^{10}}+ \frac{c^I_{14}}{L^{14}}+ \frac{c^I_{18}}{L^{18}},
\end{equation}
where $I$ stands for $D$ or $K$ and $c^I_n$ are fitting parameters effectively including all the complicated dependence on interaction matrix elements. We stress the fact that different exchange paths can give rise to perturbation terms with the same scaling dependence on the interlayer distance, that are effectively regrouped in Eq.~\ref{eq:d-dependence}. As an example, the Cr-I-Cr exchange path shares the same $L^{-14}$ dependence with a process where an electron is transferred from one Cr$-d$ to the other magnetic atom through both ligands and then transferred back via a direct $d$-$d$ process. 

Despite the underlying crude approximations, Eq.~\ref{eq:d-dependence} captures surprisingly well the evolution of both DMI and KI as a function of the overall interlayer distance, as shown in Fig. \ref{fig.DMI_KI}b. Remarkably, the non-monotonic dependence of KI can be quite naturally interpreted as arising from the competition of different super-superexchange processes occurring across the VdW gap. Notwithstanding the phenomenological nature of the fitting parameters, some general trends can be deduced. For both interaction terms, $c^I_{14}$ is negative while  $c^I_{10}$ and $c^I_{18}$ are positive: the largest coefficients are those involving $p$-$p$ hopping terms, which dominate at short separation, while in the opposite limit the first term -- showing a slower decay -- prevails. For instance, at the optimal distance $L_{\rm opt}=3.48~$\AA, one has $D(L_{\rm opt})=(366-143+20)$~$\mu$eV and $K(L_{\rm opt})=(50-25+3)$~$\mu$eV, with the dominant contribution arising from the slow-decaying $c^I_{10} {L^{-10}}$ term: a compression of 18\% causes an increase of both interactions by more than 300\% and 140\%, respectively, but the first two terms in Eq.~\ref{eq:d-dependence} largely compensate each other and the third term starts to kick in, as $D(0.82 L_{\rm opt})=(2.66-2.30+0.70)$~meV and  $K(0.82 L_{\rm opt})=(0.36-0.41+0.12)$~meV. 

It is worth reminding that the intralayer Kitaev interaction in monolayer CrI$_3$ (not to be confused with $K$ as defined in Subsection~\ref{subsec.model}) has been shown to scale quadratically with ligand iodine SOC, the transition-metal SOC contribution being negligible\cite{Xu2018Nov}. The superexchange mechanisms leading to anisotropic exchange interactions are therefore different from those discussed in the seminal Moriya's paper\cite{Moriya1960Oct}, where it was assumed that the spin of the magnetic atom couples to its own orbital moment. On the other hand, the microscopic mechanisms at play are analogous to those analysed for related transition-metal dihalides, sharing with trihalides  similar local bonding environments and strong magnetic anisotropies\cite{amoroso2020spontaneous,riedl2022}. To shed light on SOC dependence of interlayer anisotropic exchange couplings in bilayer CrI$_3$, we varied the SOC constant from $\lambda = 0$ (no SOC) to $\lambda = 2 \lambda_0$ (double the original value), as shown in the inset of Fig. \ref{fig.DMI_KI}b, and fitted the values to the power function $g(\lambda) = g_0 \lambda^n$. The fit yielded $n_D = 1.03$ and $n_K = 2.19$ (see Fig.~\ref{fig.DMI_KI}c), meaning that $D$ is a linear function of $\lambda$ whereas $K$ has nearly quadratic behavior  as the intralayer Kitaev interaction. The scaling analysis on both SOC and interlayer distance suggests that superexchange mechanisms are effective in CrI$_3$ bilayer despite the presence of a VdW gap.

To examine the influence of a stacking sequence we gradually translated the upper layer along the ${\bf a}_2$ lattice vector, starting from the LT and ending with the HT structure (Fig.~\ref{fig.DMI_KI}c). The different behavior of $D$ and $K$ with stacking alteration demonstrates the importance of angles in the Cr-I-I-Cr bonds for hopping processes that are governing the interlayer interaction. Along the inspected direction of translation, $K$ reaches it's maximum of $33 \, \mu{\rm eV}$ on the halfway between the LT and HT, whereas the maximum of $D$ of $236 \, \mu{\rm eV}$ is in the HT structure. It is interesting to see that for structures that are intermediate between LT and HT the DMI and KI are the same order of magnitude. Note that in order to find the global maximum of $D$ and $K$ one should inspect all possible directions, which is a demanding computational task that goes out of the scope of the present work.

\section{Discussion} 
\label{sec.conclusion}
\noindent
DFT calculations supported by Hamiltonian modeling reveal strong interlayer DMI in $A_1 A_2$ ($B_1 B_2$) sublattice of HT bilayer CrI$_3$. At the microscopic scale, DMI in HT bilayer emerges between the nearest and the next-nearest interlayer Cr neighbors due to broken local spatial inversion. However, due to the global C$_2$ symmetry of sublattices, the contributions from $A_1 A_2$ and $B_1 B_2$ cancel each other resulting in zero net macroscopic DMI. 
In addition to DMI, there is an order of magnitude weaker interlayer KI that does not vanish due to symmetry. In LT structure there is no DMI as the Cr-Cr pair with their surrounding iodines are centrosymmetric. Despite the DMI in HT structure dies out at the macroscopic scale, the main result of this work is the demonstration of the ability of iodine ligands to efficiently mediate the anisotropic exchange between the magnetic layers. This ability comes from the strong SOC of spatially extended I-$5p$ orbitals and is impaired if iodine atoms are replaced by other ligands that have weaker SOC. 
The developed computational procedure offers unprecedented accuracy in calculating 
the anisotropic exchange interactions. Being quite general, it can be applied in any magnetic material provided that the use of the magnetic force theorem is physically sound. The symmetry analysis and the benchmark with the reference four-state method firmly support the accuracy offered by our method.  
In addition to  the detailed analysis of the anisotropic interlayer exchange, we demonstrated that SIA heavily depends on stacking. Therefore, in estimating the Curie temperature of bilayer or bulk VdW magnet, one should not use the SIA calculated for a monolayer but instead should calculate the SIA for the system of interest.

With all being said, bilayer CrI$_3$ is not an appropriate 2D magnetic system for the experimental demonstration of interlayer DMI. Notwithstanding, we identified all the bricks needed to build one. 
%One way to continue from here is to modify the bilayer CrI$_3$ by depositing it on a proper substrate, 
%by exposing it to external electric field, or by selectively adsorbing a specie that can alternate MAE 
%from easy axis to easy plane in one of its layers. 
Instead of attempting to modify the bilayer CrI$_3$, the approach that seems more promising is to build from scratch a new heterostructure by finding an appropriate 2D magnet that can complement a layer of CrI$_3$. Using a different 2D magnet as second layer has a huge advantage in inducing DMI, as one doesn't need to worry about the spatial-inversion symmetry which is trivially broken by different chemical composition of the two layers. To efficiently mediate the DMI between the magnetic ions, a candidate 2D magnet should have ligands that feature strong SOC. Most importantly, contrary to CrI$_3$ it should show in-plane magnetic anisotropy in order to maximize the $|{\bf M}_1 \times {\bf M}_2|$ product and its MAE should be strong enough to compete with the interlayer Heisenberg exchange that favors the (anti)parallel spin configuration. 
%Otherwise, its spins will be redirected out of plane upon merging the 2D magnet 
%with CrI$_3$ and consequently the DMI will disappear.
 
Within the CrX$_3$ family of 2D magnets only CrCl$_3$ shows an easy--plane MAE\cite{Lu2020Sep}, but the other properties discredit it from potential candidacy. 
First, its MAE is extremely small and when combined with CrI$_3$ all the chances are that the interlayer Heisenberg exchange will prevail, directing CrCl$_3$ magnetization out of plane. Second, the Cr-I-Cl-Cr interaction path is far less efficient than Cr-I-I-Cr in mediating DMI, due to small SOC on Cl. 
%We partly confirmed this claim by ligand replacement in Subsection~\ref{subsec.clusters}, where we showed that replacing I with Cl reduces DMI by an order of magnitude. 
We further note that due to the lattice constant mismatch between CrI$_3$ and CrCl$_3$ one would need to match $6 \times 6$ structure of CrI$_3$ with $7 \times 7$ structure of CrCl$_3$ to build a CrI$_3$/CrCl$_3$ heterostructure, ending up with a 680 atoms in the supercell. Hence, due to high computational demands, we were not able to check these assumptions, thus leaving the search for a potential candidate for future studies.

\section{Methods}
\label{sec.methods}
DFT calculations are performed using the {\sc Vasp} code\cite{Kresse1996Oct}.
To describe the effects of electronic exchange and correlation we used the 
Perdew--Burke--Ernzerhof (PBE) functional \cite{PBE}. We did not employ and effective 
Hubbard parameter $U$ but we are aware that the choice of $U$ may affect 
the exchange coupling, as it is the case for monolayer CrI$_3$\cite{Xu2018Nov}.
The lattice constant $a = 7.005 \, \angstrom$ of monolayer CrI$_3$ is obtained from 
spin-polarized collinear DFT calculations assuming the FM ground state. 
The interlayer distance is set to $L_{\rm opt} = 3.48 \, \angstrom$, which corresponds 
to the experimental interlayer distance in bulk HT structure \cite{McGuire2015Jan}. 
The bilayer made by stacking two monolayers was not relaxed any further. 
The lattice vector along the $c$-axis was set to 30{\AA} so that the vacuum between 
periodic replicas along $c$-axis is 20 {\AA} thick. 
%We did not use the Hubbard U correction or van der Waals dispersion correction 
%to the exchange-correlation functional. 
A cutoff of $450 \, {\rm eV}$ is imposed onto the plane wave basis set 
and the total energies are converged to the precision of $10^{-9} \, {\rm eV/electron}$.
The Brillouin zone of the $2 \times 2 \, (3 \times 3)$ supercell was sampled by 
$\Gamma$-centered $4 \times 4 \times 1 \, (3 \times 3 \times 1)$ $k$-points mesh.
The results didn't change with further increase in $k$-point density owing to semiconducting 
nature of CrI$_3$. The directions of magnetic moments on Cr atoms were constrained 
using the approach exposed in Ref.\cite{Ma2015Feb}. We carefully checked whether 
the size of the sphere (RWIGS) used for calculating the magnetic moments on Cr atoms 
and the weight of the penalty functional (LAMBDA) affect the obtained $\mathcal{J}$-matrix elements.
In the end we used ${\rm RWIGS} = 1.323 \, {\angstrom}$ and ${\rm LAMBDA} = 10$. 

\begin{comment}
\section{Data availability}
Computational data used for analysis during this study are included in this published article
and its Supplementary Information. 
\end{comment}

\section*{Acknowledgment}
\noindent
The authors acknowledge support from the Italian Ministry for Research and Education through PRIN-2017 projects ``TWEET: Towards ferroelectricity
in two dimensions" (IT-MIUR Grant No. 2017YCTB59) and ``Quantum 2D: Tuning and understanding Quantum
phases in 2D materials” (IT-MIUR Grant
No. 2017Z8TS5B). SS thanks Marko Milivojevi\'{c} for very useful discussions about the role 
of spin-orbit coupling in the superexchange interaction. 

\section*{References}
%
%\bibliography{references.bib} % run this line to create .bbl file

%\addbibresource{srdjan_references.bib}
%\bibliographystyle{unsrt}
%\begin{thebibliography}{40}
%\end{thebibliography}

\end{document}

\beginsupplement

\title{SUPPLEMENTARY INFORMATION\\ 
\textcolor{white}{space} \\
Delving into the anisotropic interlayer \\ exchange in bilayer CrI$_3$
}
\date{\today}

%\pacs{}

\maketitle

\section{Benchmark of the computational method} 
Here we present the benchmark of our method of rotating perpendicular spins and compare it to the reference four-state method. We stress that our method largely reduces the number of SCF calculations that are needed for obtaining the off-diagonal $\mathcal{J}$-matrix elements. The reason is that, once the electron density is well converged in a noncollinear SCF calculation, the non-SCF calculations for different directions of the spin quantization don't cost much, as each of them takes just a few more iterations. Note that the noncollinear SCF calculations are in general tricky and one should take the opportunity to avoid performing them whenever possible. To demonstrate how much time is saved without loss in accuracy, we calculated the $\mathcal{J}_1$-matrix and SIA of monolayer CrI$_3$. We used the computational parameters reported in the Methods of the Main Text. The calculations were performed using the OpenACC GPU port of VASP 6.3.2 on a single node equipped with 4 NVIDIA Volta V100 GPUs and 256 GB of RAM. The results summarized in Table~\ref{tab.benchmark} show that 
our method can halve the computational time.  

\begin{table}[h]
	\caption{Out-of-plane SIA ($A_\perp$) and matrix elements of $\mathcal{J}_1$ obtained with the reference four-state method and with our method. As our method doesn't provide the diagonal matrix elements, in accounting the CPU time spent with four-state method we present only the time spent for calculating the off-diagonal matrix elements.}
	\begin{tabular}{ l||c|c|c c c|c|c|} 
		\hline
		method     & $A_\perp (\mu {\rm eV})$ & CPU time(s) & & $\mathcal{J}_1 (\mu {\rm eV})$ & &  CPU time(s)         \\ \hline
		           &                          &                               &   -3299   &     -6    &     0    &                                         \\ 
		four-state &  297                     & 5284        &      3   &    -2382  &   389    & 34681               \\
				   &                          &                               &       0   &      390  & -2743    &                                       \\ \hline
 		           &                          &                               &     ***   &       0   &     0    &                                       \\ 
 		our method &  265                     & 1878        &       0   &    ***    &   420    & 16554                \\
 		           &                          & 35\%        &       0   &      420  &   ***    & 48\%                 \\
		\hline
	\end{tabular}
	\label{tab.benchmark}
\end{table}

\section{Extended Moriya's rules}

 Here we extend the well-known Moriya's rules, originally devised to deduce the allowed direction of the Dzyaloshinskii vector $\bm D$ from the crystal symmetry, to the full anisotropic part of the exchange tensor, including both the antisymmetric and the symmetric components. In his seminal paper Ref. \onlinecite{Moriya1960Oct}, Moriya derived five rules for the bilinear exchange coupling between two ions located at points $A$ and $B$, where the point bisecting the straight line $AB$ (i.e., the bond vector) is denoted by $C$. Five classes of symmetry were considered, namely:
 \begin{description}
 \item[{\it a.}] inversion center at $C$
 \item[{\it b.}] mirror plane perpendicular to $AB$ and passing through $C$
 \item[{\it c.}] mirror plane including $A$ and $B$
 \item[{\it d.}] two-fold rotation axis perpendicular to $AB$ and passing through $C$
 \item[{\it e.}] $n$-fold rotation axis ($n\geq 2$) parallel to $AB$
 \end{description}
 Each of the considered bond symmetries must leave invariant the following interaction term
 \begin{equation}
 H_{AB} = \bm S_A \cdot \mathcal{J}^{AB}\cdot \bm S_B \equiv \sum_{\alpha\beta} \mathcal{J}_{\alpha\beta}^{AB} S_{A,\alpha} S_{B,\beta}
 \end{equation}
 where $\alpha,\beta$ denote cartesian components. We choose a cartesian coordinate system where the $x$ axis is parallel to the bond $AB$, and we denote the five symmetry classes above as $i$ (case {\it a}), $m_{100}$ (case {\it b}), $m_{001}$ or  $m_{010}$ (case {\it c}), $C_{2z}$ or $C_{2y}$ (case {\it d}) and $C_{nx}$ (case {\it e}). The transformation table for the cartesian components of the axial vectors $\bm S_A, \bm S_B$ is given as Table \ref{table:moriya}. We notice that inversion simply acts on the site indices, swapping A and B spins, while $m_{001}$ (or $m_{010}$) and $C_{nx}$ act on cartesian components withouth swapping site indices.
 \begin{table}
 \begin{tabular}{c|ccccc|}
 		 & $i$	 	& $m_{100}$	& $m_{001}$	& $C_{2z}$	& $C_{2x}$	\\
  \hline
 $S_{A,x}$ 	& $S_{B,x}$ 	& $S_{B,x}$	& $-S_{A,x}$	& $-S_{B,x}$	& $S_{A,x}$\\
 $S_{A,y}$ 	& $S_{B,y}$ 	& $-S_{B,y}$	& $-S_{A,y}$	& $-S_{B,y}$	& $-S_{A,y}$\\
 $S_{A,z}$ 	& $S_{B,z}$ 	& $-S_{B,z}$	& $S_{A,z}$	& $S_{B,z}$	& $-S_{A,z}$\\
 \hline
 $S_{B,x}$ 	& $S_{A,x}$ 	& $S_{A,x}$	& $-S_{B,x}$	& $-S_{A,x}$	& $S_{B,x}$\\
 $S_{B,y}$ 	& $S_{A,y}$ 	& $-S_{A,y}$	& $-S_{B,y}$	& $-S_{A,y}$	& $-S_{B,y}$\\
 $S_{B,z}$ 	& $S_{A,z}$ 	& $-S_{A,z}$	& $S_{B,z}$	& $S_{A,z}$	& $-S_{B,z}$\\
 \hline
 \end{tabular}
 \caption{Transformation table for axial vectors $\bm S_A, \bm S_B$ located at points $A$ and $B$ under point-group symmetries for the fixed point $C$ bisecting the straight line $AB$. For cases {\it c} and {\it d} we choose $m_{001}$ and $C_{2z}$ symmetry elements without loss of generality (equivalent subcases can be obtained by rotating the reference frame around the $x$ axis). For case {\it e}, we take $n=2$ for the sake of simplicity, but the result can be easily generalized.}\label{table:moriya}
 %\end{table}
 \end{table}
 We focus only on symmetry constraints for off-diagonal components of the tensor $\mathcal{J}^{AB}$. Indeed, each diagonal component $\mathcal{J}^{AB}_{\alpha\alpha}$ always complies with the considered symmetries, as these do not mix cartesian components, as clearly seen from Table \ref{table:moriya}.

 \paragraph{Inversion center at C.} Applying the transformation rules of Table \ref{table:moriya}, it is found that $S_{A,\alpha} S_{B,\beta}\mapsto S_{B,\alpha} S_{A,\beta}$. It follows that, to keep the interaction term invariant, $\mathcal{J}_{\alpha\beta}^{AB} = \mathcal{J}_{\beta\alpha}^{AB}$. Consistently with the first Moriya's rule, the antisymmetric component (i.e., the DMI) must vanish in the presence of inversion symmetry, and the anisotropic part of the full exchange tensor must be symmetric, with no further constraints on off-diagonal components of the pseudo-dipolar term. Using the Dzyaloshinskii vector and the analogous vector $\bm K$ we introduced to denote the off-diagonal components of the symmetric anisotropic tensor, inversion implies that $\bm D=0$ but $\bm K \neq 0$.

 \paragraph{Mirror plane perpendicular to AB and passing through C.} In this case one finds:
 \begin{eqnarray}
 S_{A,x} S_{B,y} &\mapsto& - S_{B,x} S_{A,y}, \nonumber\\
 S_{A,x} S_{B,z} &\mapsto& - S_{B,x} S_{A,z}, \nonumber\\
 S_{A,y} S_{B,z} &\mapsto& + S_{B,y} S_{A,z},\nonumber
 \end{eqnarray}
 implying $\mathcal{J}_{xy}^{AB} = -\mathcal{J}_{yx}^{AB}$, $\mathcal{J}_{xz}^{AB} = -\mathcal{J}_{zx}^{AB}$ and $\mathcal{J}_{yz}^{AB} = +\mathcal{J}_{zy}^{AB}$. The mirror symmetry implies that two purely antisymmetric components exist, corresponding to a Dzyaloshinskii vector $\bm D$ perpendicular to the AB ($x$) axis, or equivalently lying in the mirror plane, as stated by the second Moriya's rule. Additionally, there is a purely symmetric part of the anisotropic exchange tensor coupling spin components lying in the plane perpendicular to the AB ($x$) axis.

 \paragraph{Mirror plane including A and B.} In this situation, the reflection does not swap sites and, choosing the mirror plane to be perpendicular to the axis $z$ (and parallel to the axis $x$), one gets:
 \begin{eqnarray}
 S_{A,x} S_{B,y} &\mapsto& + S_{A,x} S_{B,y}, \nonumber\\
 S_{A,x} S_{B,z} &\mapsto& - S_{A,x} S_{B,z}, \nonumber\\
 S_{A,y} S_{B,z} &\mapsto& - S_{A,y} S_{B,z}.\nonumber
 \end{eqnarray}
 It follows that $\mathcal{J}_{xz}^{AB} = -\mathcal{J}_{xz}^{AB}\equiv 0$ and $\mathcal{J}_{yz}^{AB} = -\mathcal{J}_{yz}^{AB}\equiv 0$, while no symmetry constraints act on the component $\mathcal{J}_{xy}^{AB}$. The latter condition is compatible with the third Moriya's rule, stating that $\bm D$ is perpendicular to the mirror plane $xy$, but a symmetric anisotropic part is also allowed for spin components lying within the same reflection plane.

 \paragraph{Two-fold rotation axis perpendicular to AB and passing through C.} Transformation rules imply that, for the two-fold rotation axis parallel to the axis $z$ ($\perp x$):
 \begin{eqnarray}
 S_{A,x} S_{B,y} &\mapsto& + S_{B,x} S_{A,y}, \nonumber\\
 S_{A,x} S_{B,z} &\mapsto& - S_{B,x} S_{A,z}, \nonumber\\
 S_{A,y} S_{B,z} &\mapsto& - S_{B,y} S_{A,z}.\nonumber
 \end{eqnarray}
 It follows that the anisotropic exchange tensor can be decomposed in a purely antisymmetric part with  $\bm D$ perpendicular to the two-fold axis, as $\mathcal{J}_{xz}^{AB} = -\mathcal{J}_{zx}^{AB}$ and $\mathcal{J}_{yz}^{AB} = -\mathcal{J}_{zy}^{AB}$ (consistenly with Moriya's fourth rule), as well as a purely symmetric part for spin components lying in a plane perpendicular to the rotation axis.

 \paragraph{Two-fold rotation axis parallel to AB.} As for the case {\it c}, any rotation around an axis parallel to the bond $AB$ does not swap sites, and for a two-fold rotation one gets:
 \begin{eqnarray}
 S_{A,x} S_{B,y} &\mapsto& - S_{A,x} S_{B,y}, \nonumber\\
 S_{A,x} S_{B,z} &\mapsto& - S_{A,x} S_{B,z}, \nonumber\\
 S_{A,y} S_{B,z} &\mapsto& + S_{A,y} S_{B,z}.\nonumber
 \end{eqnarray}
 It follows that $\mathcal{J}_{xy}^{AB} = -\mathcal{J}_{xy}^{AB}\equiv 0$ and $\mathcal{J}_{xz}^{AB} = -\mathcal{J}_{xz}^{AB}\equiv 0$, and no symmetry constraint exists on the $\mathcal{J}_{yz}^{AB}$ component. Its antisymmetric part will give rise to a Dzyaloshinskii vector parallel to the bond axis $x$ (as stated by the fifth Moriya's rule), while the symmetric part will couple spin components lying in a plane perpendicular to it.

 The Moriya's rules extended also to the symmetric part of the anisotropic exchange tensor are summarized in Table \ref{table:moriya2}, in terms of the Dzyaloshinskii vector $\bm D$ and the $\bm K$ vector introduced in the main text.
 \begin{table}
 \begin{tabular}{p{0.5cm}|>{\centering\arraybackslash}p{2.8cm}%
 >{\centering\arraybackslash}p{2.8cm}%
 >{\centering\arraybackslash}p{2.8cm}%
 >{\centering\arraybackslash}p{2.8cm}%
 >{\centering\arraybackslash}p{2.8cm}|}
 		 & inversion	   & $m_\perp$		      & $m_\parallel$	     & $C_{2\perp}$     	     & $C_{n\parallel}$	\\[0.1cm]
 \hline
 $\bm D$         & $=0$	   & $\parallel$ mirror plane &	$\perp$ mirror plane & $\perp$ two-fold axis & $\parallel AB$\\[-0.cm]
                 &         & or $\perp AB$             &			    &			    &    \\[0.1cm]
		 \hline
 $\bm K$         & $\neq 0$ & $\parallel AB$           & $\perp$ mirror plane & $\parallel$ two-fold axis & $\parallel AB$\\[0.1cm]
 \hline
 \end{tabular}
 \caption{Generalized Moriya's rules for the anisotropic part of the exchange tensor, including both symmetric and antisymmetric components, described by the $\bm K$ and $\bm D$ vectors, respectively. All symmetry elements leave the $C$ point bisecting the $AB$ invariant, and their subscripts $\perp$ and $\parallel$ denote their corresponding orientation with respect to the $AB$ direction.}\label{table:moriya2}
 \end{table}

%\textcolor{red}{to PAOLO: please write here what you sent to me a while ago about Moriya rules} 

\section{$\mathcal{J}$-matrices of reported structures}
Here we present the $\mathcal{J}$-matrices of the structures depicted in Fig.~2 of the Main Text 
and of the $A_1 A_2$, $A_1 B_2$, $B_1 A_2$ and $B_1 B_2$ sublattices of HT bilayer CrI$_3$. 
%We calculated all the matrix elements and checked whether the obtained values comply with the extended Moriya's rules afterwards. 
$S=3/2$ is used the energy mapping analysis. All the values are given in $\mu{\rm eV}$ units.

\begin{equation}
\mathcal{J_{\rm LT1}} = 
\begin{pmatrix}
***   &  0     &   0    \\
0    & ***       &   0    \\
0    &  0     & ***       \\
\end{pmatrix}, 
\quad
\mathcal{J_{\rm LT2}} = 
\begin{pmatrix}
***&  -17   &   24   \\
-16   &*** &   -14  \\
24   &  -14   & ***\\
\end{pmatrix}, 
\end{equation}

\begin{equation}
\mathcal{J_{\rm HT1}} = 
\begin{pmatrix}
***&  -168  &  -50   \\
168   & ***&  0     \\
50   &   -1   &*** \\
\end{pmatrix}, 
\quad
\mathcal{J_{\rm HT2}} = 
\begin{pmatrix}
***&  -4     &  166   \\
4    &***  &  -25   \\
-166   &  -25    &*** \\
\end{pmatrix}, 
\end{equation}

For HT structure the $\mathcal{J}^\updownarrow$ tensor and its decomposition onto $A$ and $B$ sublattices
\begin{equation}
\mathcal{J^\updownarrow} = 
\begin{pmatrix}
***   &  0     &   0    \\
0    & ***       &  -18    \\
0    &  -18     & ***       \\
\end{pmatrix}, 
\end{equation}

\begin{equation}
\mathcal{J}^\updownarrow_{A_1A_2} = 
\begin{pmatrix}
***&  -139  &  -190  \\
139   & ***&  -29   \\
190  &   -29  &*** \\
\end{pmatrix}, 
\quad
\mathcal{J}^\updownarrow_{A_1B_2} = 
\begin{pmatrix}
***    &  -50    &  -34   \\
-50  & ***  &  2     \\
-34    &   2     &*** \\
\end{pmatrix}, 
\end{equation}

\begin{equation}
\mathcal{J}^\updownarrow_{B_1A_2} = 
\begin{pmatrix}
***&   50   &   34   \\
50   & ***&  2     \\
34  &    2   &*** \\
\end{pmatrix}, 
\quad
\mathcal{J}^\updownarrow_{B_1B_2} = 
\begin{pmatrix}
***&  139    &  191   \\
-139 &***  &  -29   \\
-190   &  -29    &*** \\
\end{pmatrix}, 
\end{equation}

\bibliography{../srdjan_references.bib}